# Electron Transfer on Impurity doped Graphene Nanoribbon

Hiroyoshi Tsuyuki, Shoichi Sakamoto and Mitsuyoshi Tomiya

*Department of Materials and Life Science, Seikei University,*

*Kichijyoji-Kitamachi 3-3-1, Musashino-shi, Tokyo, 180-8633, JAPAN*

dd106102@cc.seikei.ac.jp

**Abstract:** Electronic transport properties in armchair shaped edges graphene nanoribbons (AGNRs) doped various impurities have been simulated by the non-equilibrium Green's function approach combined with the first principle calculation based on the density functional theory. We have observed that impurity levels appear in electronic structures, and that the quantization of transmission function is moderated for doped AGNRs. The I-V characteristic can be computed from the transmission function. Our simulation results show that AGNRs doped impurities have higher conductance than the non-doped one.

## 1. INTRODUCTION

Recently, a graphene has been extensively studied due to its unique physical properties originated from the unique two-dimensional (2D) honeycomb lattice structure [1]. Graphene nanoribbons (GNRs) are narrow strips of a single layer graphene with nanometer sized width and armchair or zigzag shaped edges.

A graphene sheet is half-metallic, whereas GNRs have ether metallic or semiconducting feature by the shape of edge [1-3], which is however different from that in the carbon nanotube (CNT) [4]. Electronic properties of the GNR are controlled by the width and the edge shape [2, 3]. A graphene sheet does not have a bandgap so that a conduction band and a valence band touch at K-point. On the other hand, the GNR with armchair shaped edges (AGNR) which has semiconducting feature has a bandgap by the quantum confinement effect to the width direction. The bandgap in the AGNR can be controlled by the ribbon width [3].

To apply for various devices, it is important to dope impurities on semiconductor. Boron (B) and nitrogen (N) atoms are typical dopants. It is expected to have various functions and interesting physical properties by the doping chemically for new device applications using carbon based materials. For carbon materials such as the GNR and the CNT, electronic properties, e.g., the electronic structure, the conductance etc., in doping impurities are now being studied intensively [5-8]. For example, B atom as a substitutional atom is often considered for p-type doping, N atom is also adopted as n-type doping. We are interested in effects of dopants to the electronic transfer on GNRs.

In this work, we focus on the AGNR. We have numerically investigated electronic transport properties on AGNRs doped some of atoms and molecules on the surfaces by the non-equilibrium Green's function (NEGF) formalism [9] combined with the standard first principle calculation. Electronic structures of doped AGNRs have first been computed by the first principle calculation. For electronic transport, I-V characteristics are finally obtained by calculating the transmission function considering to apply the bias voltage.

## 2. METHOD

Our calculation for the electronic structure is done by the standard first principle method, which is based on the density functional theory (DFT). We have used the DFT program package SIESTA. The pseudo potential for the Kleinman-Bylander (KB) potential, and the generalized gradient approximation (GGA) for the exchange correlation term is adopted. For our simulation, we take following conditions; the k-point sampling is 256 points, and the energy cutoff is 350 Rydbergs.

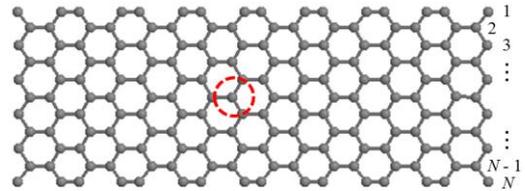

**Figure 1:** The carbon atomic structure of AGNR modeling as the super cell. $N$ is the number of dimer lines, $N = 11$, across the ribbon width. At the center of the ribbon surface, as shown dashed line circle, some impurity has been attached on site. This structure is relaxed by the CG optimization.

We have made the AGNR structure in which the number of carbon atoms of the AGNR is 176 atoms, and the bond length between carbons is 1.48 Angstroms, as shown Figure 1. This bond length is obtained by the optimization of the structure with the conjugate gradient (CG) method. In the DFT calculation, the bond length is slightly longer.

We have also considered connections to electrodes which was used same type GNRs. Then the quantum transmission is computed by the NEGF theory combined with the first principle calculation using the TRANSIESTA package. For the NEGF formalism, the three parts; the left electrode, the contacting part, and the right electrode, are required.

In the NEGF approach, the Green's function at energy $E$ is calculated as

$$G(E) = (E - H)^{-1}, \quad (1)$$

after the Hamiltonian $H$ is obtained by the DFT calculation with SIESTA. Considering the non-equilibrium situation, the bias voltage is defined as the difference of chemical potentials $\mu_L$ and $\mu_R$ in the left and right electrodes, $eV = \mu_L - \mu_R$. The transmission is calculated as

$$\begin{aligned}T(E) &= tr[\mathbf{t}^{\dagger}\mathbf{t}](E) \\ &= tr[\mathbf{\Gamma}_L(E)\mathbf{G}^{\dagger}(E)\mathbf{\Gamma}_R(E)\mathbf{G}(E)]\end{aligned}, \quad (2)$$

with the transmission amplitude matrix,

$$\mathbf{t}(E) = \mathbf{\Gamma}_R(E)^{1/2}\mathbf{G}(E)\mathbf{\Gamma}_L(E)^{1/2}, \quad (3)$$

where,

$$\mathbf{\Gamma}_{L,R} \equiv i[\mathbf{\Sigma}_{L,R}(E) - \mathbf{\Sigma}^{\dagger}_{L,R}(E)]/2, \quad (4)$$

and $\mathbf{\Sigma}_{L,R}$ is the self-energy in the left or right electrode. By the Landauer formulation, the current is calculated from the transmission function,

$$I = \frac{2e^2}{h}\int_{\mu_L}^{\mu_R}d\varepsilon[n_F(\varepsilon - \mu_L) - n_F(\varepsilon - \mu_R)]T(\varepsilon), \quad (5)$$

where $n_F$ is the Fermi distribution function.

## 3. RESULTS

Four kinds of impurities; hydrogen (H) atom, hydroxide (OH), methyl ($CH_3$), and sodium (Na) atom are considered here. Just one atom or molecule is attached on (not substituted) the site of the AGNR surface (as shown Figure 1). The distance between the carbon atom of the AGNR surface and attached atoms or molecules is set 1.0 Angstroms before the structures are relaxed by the CG method. Then the total energy of the system becomes the local minimum.

First, the electronic structure of the non-doped AGNR has been simulated. The Fermi energy $E_F$ is -6.105eV, and the bandgap, which is about 0.4eV, exists at K-point in this semiconducting AGNR (Figure 2(a)). When the H atom and $CH_3$ are attached on the AGNR surface, these energy levels of impurity states appear in the AGNR bandgap (Figure 2(b), (c)). These levels are equivalent to the Fermi-energy; $E_F$ = -5.965eV for H doped and $E_F$ = -5.989eV for $CH_3$ doped, which are slightly higher than that of the non-doped AGNR. Band structures of other impurities; OH and Na, doped AGNRs are also shows in Figure 2(c)-(e). In case of OH impurity, the impurity level also appears in the bandgap of the AGNR. Fermi energy, $E_F$ = -6.215eV, is slightly lower than the non-doped AGNR. On the contrary, the Na impurity level appears in the conduction band, $E_F$ = -5.252eV, thus the Na-doped AGNR is no longer the semiconductor.

In order to research properties of electronic transport for AGNRs doped impurities, transmission functions have been calculated. It is shown in Figure 3 for the zero-bias situation. For the non-doped AGNR, it is confirmed the transmission function is quantized against energies (Figure 3(a)). In contrast, the quantization of transmission functions are moderated for doped AGNRs (Figure 3(b)-(e)).

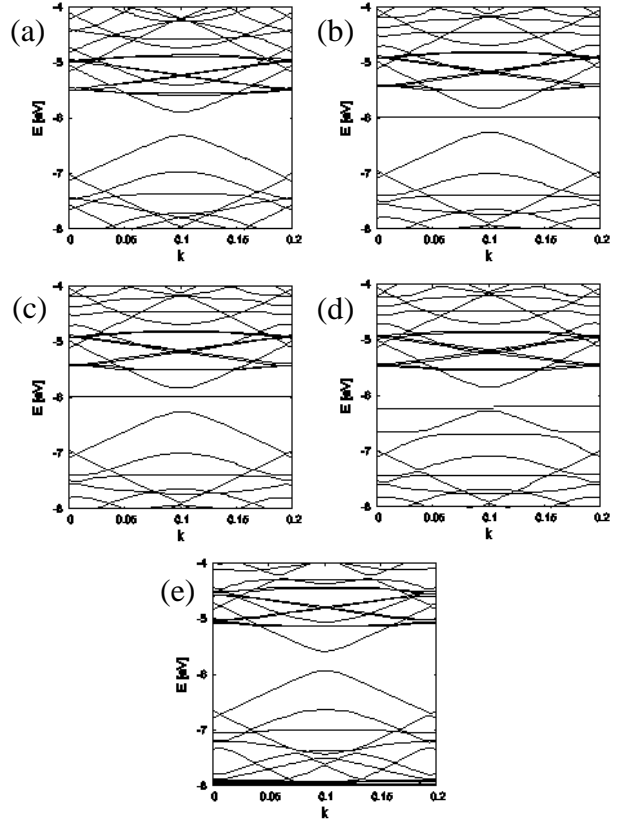

**Figure 2:** Band structures for (a) non-doped, (b) H-atom, (c) CH3, (d) OH, and (e) Na-atom, doped AGNRs. At K-point, the energy gap exist between the conduction and the valence band.

In cases of H atom, OH, and $CH_3$ impurities doping, the quantization of transmission functions is moderated. On the DOS, the contribution of the impurity level is clearly observed in the bandgap for OH doped AGNR. It is observed that dips appear on the transmission function, where impurity states exist (Figure 3(d)) [7]. However, contributions of H atom and $CH_3$ impurity levels are almost not observed (Figure 3(b)-(c)). For these impurities doping, no transmission exist at Fermi-levels. For Na impurity doping, the quantization of the transmission is broken (Figure 3(e)).

I-V characteristics of doped AGNRs are also simulated using transmission functions on finite biases (Figure 4). The bias voltage is applied from 0.0eV to 2.0eV. Since this AGNR is a semiconductor, the current starts to flow from about 0.4eV. It is observed that the conductance of AGNR semiconductor is increased by doping these impurities.

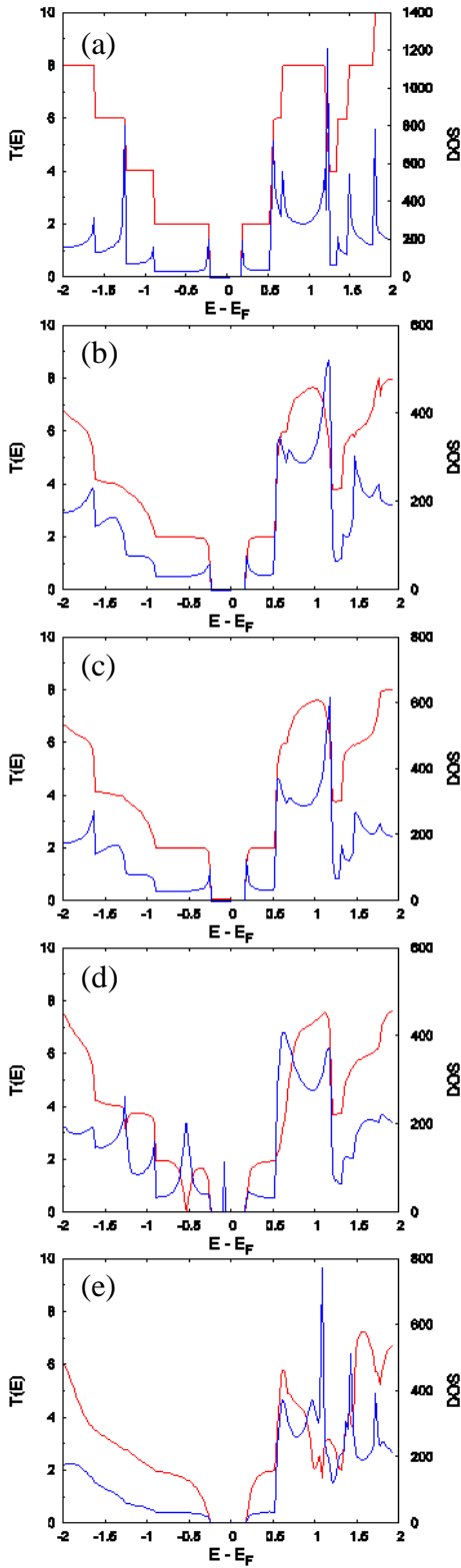

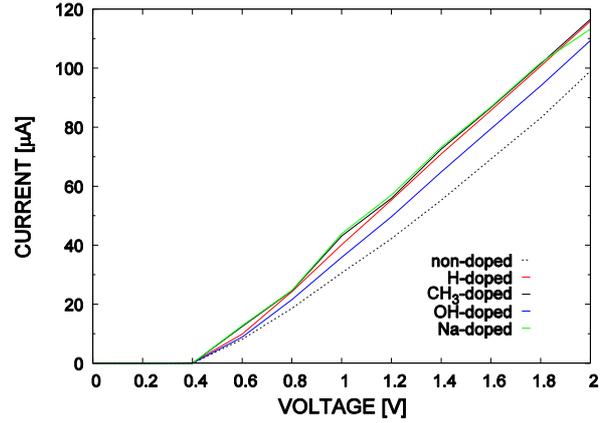

**Figure 4:** I-V characteristics of AGNRs. The conductance in doped AGNR is higher than that of non-doped AGNR denoted by the dashed line.

## 4. CONCLUSION

We have numerically investigated electronic properties of the AGNR doped impurity. When some atom or molecule is attached on the AGNR surface, impurity levels appear in bandgaps of AGNRs, whereas no impurity level of metal appears in the bandgap. It is observed that the transmission function is quantized against energies in non-doped AGNR, but the quantization is moderated in some doped AGNRs. I-V characteristics of doped AGNRs are evaluated. Especially, we have found that AGNRs doped impurities have higher conductance than the non-doped AGNR.

**Figure 3:** Transmission functions as shown the red line around the Fermi energy for (a) non-doped, (b) H-atom, (c) CH3, (d) OH, and (e) Na-atom, doped AGNRs. The blue line shows the density of states. At the Fermi energy, there exist little impurity states in doped one.